\documentclass{emulateapj}
 
\usepackage{graphicx}
\usepackage{epsfig} 
\usepackage{natbib}
\usepackage{times}
\usepackage{amssymb,amsmath}
\usepackage{color}

\newcommand{\p}[1]{{{#1}}}
 
\shortauthors{Lionello et al.}
\shorttitle{Contribution of Jets to Solar Wind}

\begin{document}

\title{The Contribution of Coronal Jets to the Solar Wind}   

\author{ R.~Lionello,$^1$ T.~T\"or\"ok,$^1$ V.~S.~Titov,$^1$ J.~E.~Leake,$^2$ Z.~Miki\'c,$^1$ J.~A.~Linker,$^1$ and M.~G.~Linton$^2$}
\affil{$^1$Predictive Science Inc., 9990 Mesa Rim Rd., Ste. 170, San Diego, CA 92121, USA; \email{lionel@predsci.com}}
\affil{$^2$US Naval Research Laboratory 4555 Overlook Ave., SW Washington, DC 20375, USA}

\begin{abstract}
Transient collimated plasma eruptions in the solar corona, commonly known as coronal (or X-ray) jets, are among the most interesting manifestations of solar activity. It has been suggested that these events contribute to the mass and energy content of the corona and solar wind, but the extent of these contributions remains uncertain. We have recently modeled the formation and evolution of coronal jets using a three-dimensional (3D) magnetohydrodynamic (MHD) code with thermodynamics in a large spherical domain that includes the solar wind. Our model is coupled to 3D MHD flux-emergence simulations, i.e, we use boundary conditions provided by such simulations to drive a time-dependent coronal evolution. The model includes parametric coronal heating, radiative losses, and thermal conduction, which enables us to simulate the dynamics and plasma properties of coronal jets in a more realistic manner than done so far. \p{Here we employ these simulations to calculate the amount of mass and energy transported by coronal jets into the outer corona and inner heliosphere. Based on observed jet-occurrence rates, we then estimate the total contribution of coronal jets to the mass and energy content of the solar wind to $(0.4-3.0)$\,\% and (0.3-1.0)\,\%, respectively. Our results are largely consistent with the few previous rough estimates obtained from observations, supporting the conjecture that coronal jets provide only a small amount of mass and energy to the solar wind. We emphasize, however, that more advanced observations and simulations are needed to substantiate this conjecture.}
\end{abstract}


\section{Introduction}
\label{s:intro}
%
Coronal (or X-ray) jets are collimated, transient eruptions of hot plasma observed predominantly in coronal holes and at the periphery of active regions, driven by flux emergence or cancellation \citep[e.g.,][]{1992PASJ...44L.173S}. ``Standard jets'' are launched by reconnection between open magnetic fields and expanding closed fields, while the more energetic ``blowout jets'' additionally involve an eruption of the closed fields \citep{2010ApJ...720..757M}. For detailed descriptions of coronal jets see the recent reviews by \citet{2016arXiv160303258I}, \citet{2016ASPC..504...185T}, and \cite{Raouafi2016}.  

Coronagraph and in-situ observations revealed that coronal jets can produce signatures in the outer corona, sometimes even at Earth or beyond \citep[][]{1997ESASP.415..103S,1998ApJ...498L.165W,1999ApJ...523..444W,2007ApJ...659.1702C,2012ApJ...750...50N,2014SoPh..289.2633J,2014ApJ...784..166Y}. It was therefore suggested that they contribute to powering the corona and solar wind \citep[e.g.,][]{2007Sci...318.1591S,2007Sci...318.1580C}.\footnote{For simplicity, we henceforth refer to the outer corona and solar wind collectively as SW.}

Quantifying this contribution from observations is difficult, and only few attempts have been made so far. \citet{2013ApJ...776...16P} obtained a ``crude estimate'' for the energy budget of coronal jets,  which was $\sim10^4$ times smaller than required for maintaining the SW. \cite{2007Sci...318.1580C} suggested a mass-contribution of coronal jets to the SW of $\approx$10\,\%, though \cite{2015ApJ...801..124Y} remarked that this value should be 2.5 times smaller. Using heliospheric images from SMEI \citep{2004SoPh..225..177J}, \citet{2014ApJ...784..166Y} estimated mass and energy contributions of 3.2\,\% and 1.6\,\%, respectively. Finally, \citet{2015A&A...579A..96P} estimated the energy provided by coronal jets to be less than 1\,\% of what is required to heat a polar coronal hole. They emphasized, though, that many high-temperature jets may be missed by the observations \citep[][]{2015ApJ...801..124Y}. Thus, while it appears that the contribution of coronal jets to the SW may be small compared to the potential contribution by the much more numerous jet-like events on smaller scales \citep[][]{2011ApJ...731L..18M}, its exact amount remains uncertain.

Numerical models provide a powerful tool to address this question. However, MHD simulations of coronal jets \citep[e.g.,][]{1995Natur.375...42Y,1996PASJ...48..353Y,2008ApJ...683L..83N,2008ApJ...673L.211M,2009A&A...506L..45G,2009ApJ...691...61P,2010ApJ...715.1556R,2013ApJ...769L..21A,2013ApJ...771...20M,2014ApJ...789L..19F} were so far restricted to a limited treatment of the energy transfer in the corona and to heights far below the SW-acceleration region. Recently, \citet{2016ASPC..504...185T} presented MHD simulations of coronal jets that overcame both limitations. Using these simulations, we evaluate here the mass and energy contribution of coronal jets to the SW, and compare our results to previous observational estimates.

\section{MODEL AND SIMULATIONS}
\label{sec-model}
%
The simulations we use for our analysis are described in \cite{2016ASPC..504...185T}, and a more detailed account will be provided in a forthcoming publication (T\"or\"ok {\em et al.}, in preparation). They were performed using the 3D, viscous, resistive MHD model in spherical coordinates developed at Predictive Science Inc. \citep{1999PhPl....6.2217M,2009ApJ...690..902L}. The model incorporates thermal conduction parallel to the magnetic field, radiation losses, and parametric coronal heating. The nonuniform grid covers 1--20\,$R_\sun$, where $R_\sun$ is the solar radius, using $381\times341\times403$ points in $r\times\theta\times\phi$. The smallest radial grid interval at $r=R_\sun$ is $\approx320$ km; the angular resolution is $\approx0.5\arcsec$ in the area containing the jet. A purely radial coronal magnetic field, produced by a magnetic monopole located at Sun center is prescribed. Starting from an initial thermodynamic SW-solution \citep{2009ApJ...690..902L}, we relax the system to a spherically symmetric, steady-state solar wind. Afterwards, using the method of \citet{2013ApJ...777...76L}, we emerge a flux rope whose structure and evolution was previously computed with the model of \citet{2013ApJ...778...99L}. This is done by prescribing the values of the magnetic field and calculating the velocities using the characteristic equations at $r=1\,R_\sun$.

\begin{figure}[t]
\centering
\includegraphics[width=0.477\textwidth]{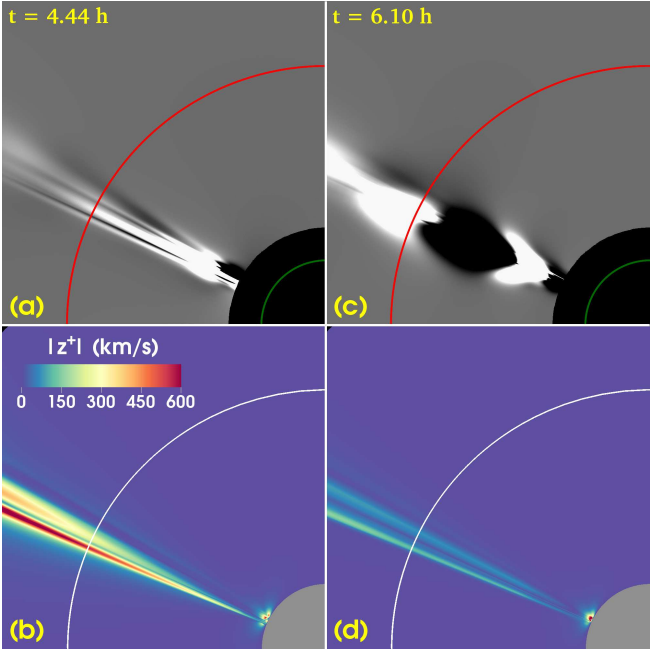}
\caption
{
(a) Synthetic running-difference image of polarization brightness
for run \emph{Jet 1}, about 2.7 (0.2) hrs after onset of the standard 
(blowout) jet. The green (red) circle marks the solar surface 
($r=4~R_\sun$).
(b) Magnitude of Els\"asser variable in a plane intersecting the 
jet; same time as in (a).
(c) Same as (a) 1.66\,hrs later, showing coronal signatures of the 
blowout jet.  
(d) Same as (b) at the time shown in (c). The times
correspond to peak energy fluxes at $r=4~R_\sun$ 
(cf. Figures\,\ref{fig-poynting_wave} and \ref{fig-f1_f2}). 
}
\label{fig-jet1}
\end{figure}

Here we analyze three runs. In all cases, the system is in a steady-state at $t=0$, when the flux emergence is started. In our main run, {\em Jet 1}, emergence is imposed for 4.5\,hrs, and the evolution is modeled for a total of 12\,hrs. The simulation produces a standard jet at $t\approx1.7$\,h, launched by the first strong occurrence of magnetic reconnection in the current layer that forms between the emerging flux and the open coronal field. Several reconnection episodes follow, driven by the continuous expansion of the emerging flux rope. At $t\approx4.2$\,h a blowout jet is produced, associated with the eruption of the rope, which again is followed by several reconnection episodes, now occurring in the current layer that forms below the eruption. The blowout jet is much stronger than the standard jet, with a peak kinetic energy of $3.6 \times 10^{28}$\,erg, compared to only $5.2 \times 10^{26}$\,erg for the standard jet.
\p{Plasma is ejected upwards by reconnection outflows during the jets, and additionally by the flux-rope eruption during the blowout jet. Both jets produce Alfv\'en waves, which travel much faster than the ejected plasma, due to the relatively large coronal Alfv\'en speed in our model (several 1000\,km\,s$^{-1}$). 

In Figure\,\ref{fig-jet1} we show manifestations of the jets in the outer corona, using synthetic difference-images of the polarization brightness \p{\citep{2011ASPC..437...99R}} and the magnitude of the Els\"asser variable, $\mathbf{z_+}= \mathbf{v_\perp} - \mathbf{B_\perp}/\sqrt{4 \pi \rho}$, which represents an outward propagating Alfv\'en perturbation along a radial magnetic field line \citep[e.g.,][]{2001ApJ...548..482D}. At $t=4.44$\,h an elongated brightness-enhancement and a clear wave-signature are visible in panels (a) and (b), respectively. The former is associated with plasma ejected by the standard jet, while the latter is associated with the Alfv\'en wave triggered by the blowout jet (the significantly weaker Alfv\'en wave triggered by the standard jet has already left the field-of-view at this time). 
%
%
At $t=6.1$\,h the plasma ejected by the blowout jet becomes visible, while there are no strong wave-signatures anymore. The brightness-image shows density-enhancements and rarefactions, reminiscent of narrow coronal mass ejections and ``streamer blobs'' \citep[e.g.,][]{1999JGR...10424739S}.    
}

Our second run, {\em Jet 2}, has the same settings as {\em Jet 1}, but here we stop the flux emergence after 2\,hrs, so that only the standard jet occurs. This run is calculated for 6\,hrs. In both runs ohmic heating is included, so that magnetic energy dissipated through the resistivity (corresponding to a resistive diffusion time $\tau_R\approx4\times 10^5$ hrs) contributes to the energy equation. Finally, in {\em Jet 3}, we again impose the emergence for\,4.5 hrs, but turn off ohmic heating. An inspection of this run reveals that it is essentially a lower-energy version of {\em Jet 1}, so we stopped the calculation shortly after the occurrence of the blowout jet, at $t=4.8$\,h.

\begin{figure}[t]
\centering
\includegraphics[width=0.477\textwidth]{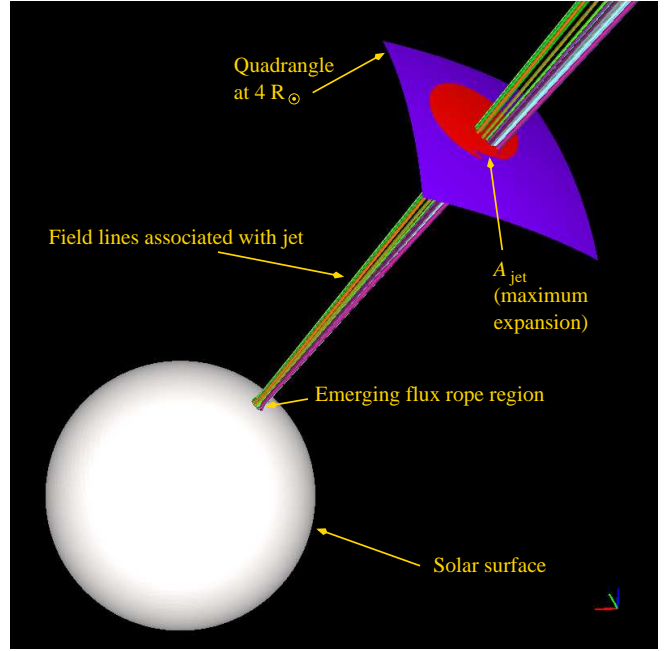}
\caption
{
To calculate energy and mass fluxes, we consider a quadrangle 
$Q$ at $r=4\,R_\sun$, $0.87\,\leq\,\theta\,\leq\,1.31$, $2.26\,\leq\,\phi\,\leq\,2.70 $, which
contains the jet at all times. For our analysis we then select an area $A_\mathrm{jet}$ that corresponds to the jet's maximum transverse expansion.
}
\label{fig-flux}
\end{figure}

\begin{figure*}[t]
\centering
\includegraphics[width=.97\textwidth]{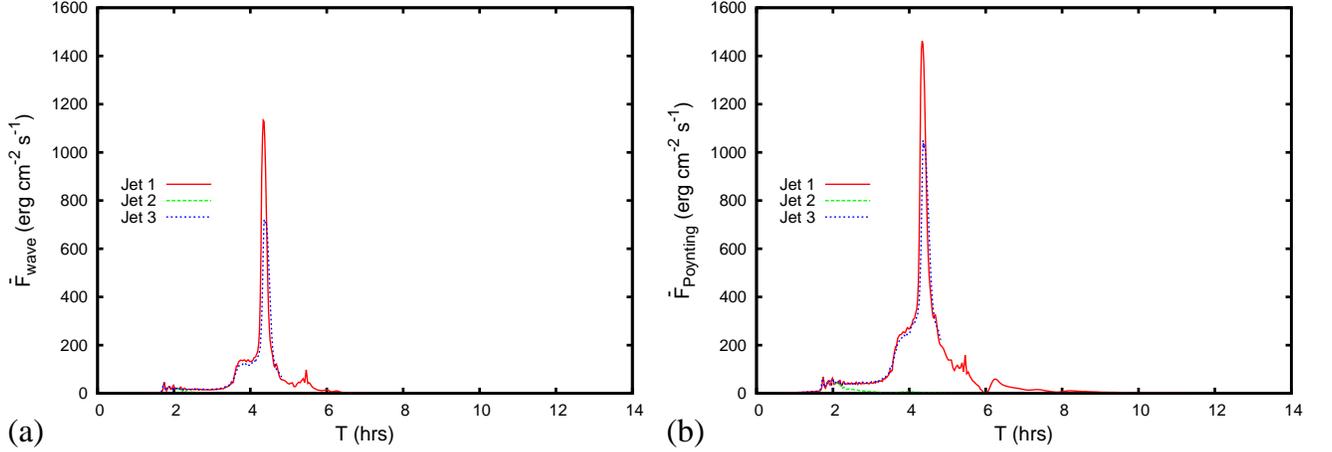}
\caption
{
(a) Average wave-energy flux at $r= 4~R_\odot$ for runs 
{\em Jet 1--3}, calculated using Equation\,(\ref{eq-flux}). 
(b) Same for average Poynting flux, using Equation\,(\ref{eq-poynting}). 
}
\label{fig-poynting_wave}
\end{figure*}

\section{ENERGY AND MASS FLUXES INTO THE SOLAR WIND}
\label{sec-fluxes}
%
Following \citet{2013ApJ...776...16P} and \citet{2015A&A...579A..96P}, we calculate the mass and energy fluxes associated with our jets, and integrate them in time and space to find the total mass and energy injected into the SW. Different from these authors, we perform our analysis high in the corona, at $r=4~R_\sun$, and use the Poynting flux, rather than the wave-energy flux, for a more accurate calculation of the total energy flux (\S\,\ref{ss:wave_vs_poynting}). For the integration we use a spherical quadrangle $Q$ at $r=4~R_\odot$ that contains the jet-signatures at all times (Figure\,\ref{fig-flux}). The area $A_\mathrm{jet}$ corresponds to the maximum lateral expansion of the blowout jet as it passes through $Q$ at $t= 6.1$\,h (see Figure~\ref{fig-jet1}(c)). We define $A_\mathrm{jet}$ as the area within which the perturbation in the sum of the kinetic and magnetic energies is larger than $3\%$ at that time. As in the above-mentioned papers, we consider the kinetic, potential, enthalpy, and wave energy fluxes:
\begin{eqnarray}
F_\mathrm{kin}= \frac{1}{2}\rho\,v^2\,v_r, &\phantom{=}&
F_\mathrm{pot}= \rho\,\frac{GM_\odot}{r}\,v_r,  \nonumber \\
F_\mathrm{enth}= \frac{\gamma}{\gamma-1}\,p\,v_r, &\phantom{=}& 
F_\mathrm{wave}= \sqrt{\frac{\rho}{4 \pi} } \,v_\perp^2\,B, 
\label{eq-flux}
\end{eqnarray}
and the radial component of the Poynting flux,
\begin{equation}
F_\mathrm{Poynting}=  \frac {c}{4 \pi}
\left.\mathbf{E}\times\mathbf{B}\right|_r. 
\label{eq-poynting}
\end{equation}
Fluxes due to radiation and thermal conduction are neglected \citep{2015A&A...579A..96P}. For each flux $\mathrm{x}$ we determine the constant background $F_\mathrm{x}^\mathrm{(bg)}$ produced by the steady-state SW (which is zero for $F_\mathrm{wave}$ and $F_\mathrm{Poynting}$). Then we calculate the net power through $Q$ due to the fluxes by integrating $F_\mathrm{x}$ and removing the background:
\begin{equation}
P_\mathrm{x}(t)=\iint\limits_Q (F_\mathrm{x}(t,\theta, \phi) -F_\mathrm{x}^\mathrm{(bg)}) \, \sin(\theta)
\, d\theta \, d \phi. 
\label{eq-pt}
\end{equation}
Dividing each $P_\mathrm{x}$ by $A_\mathrm{jet}$, we obtain the average, background-subtracted energy fluxes associated with the jets as functions of time:
\begin{equation}
\bar{F}_\mathrm{x}(t)=\frac{P_\mathrm{x}(t)}{A_\mathrm{jet}}.
\label{eq-fbar}
\end{equation}
%

\begin{figure}
\centering
\includegraphics[width=.5\textwidth]{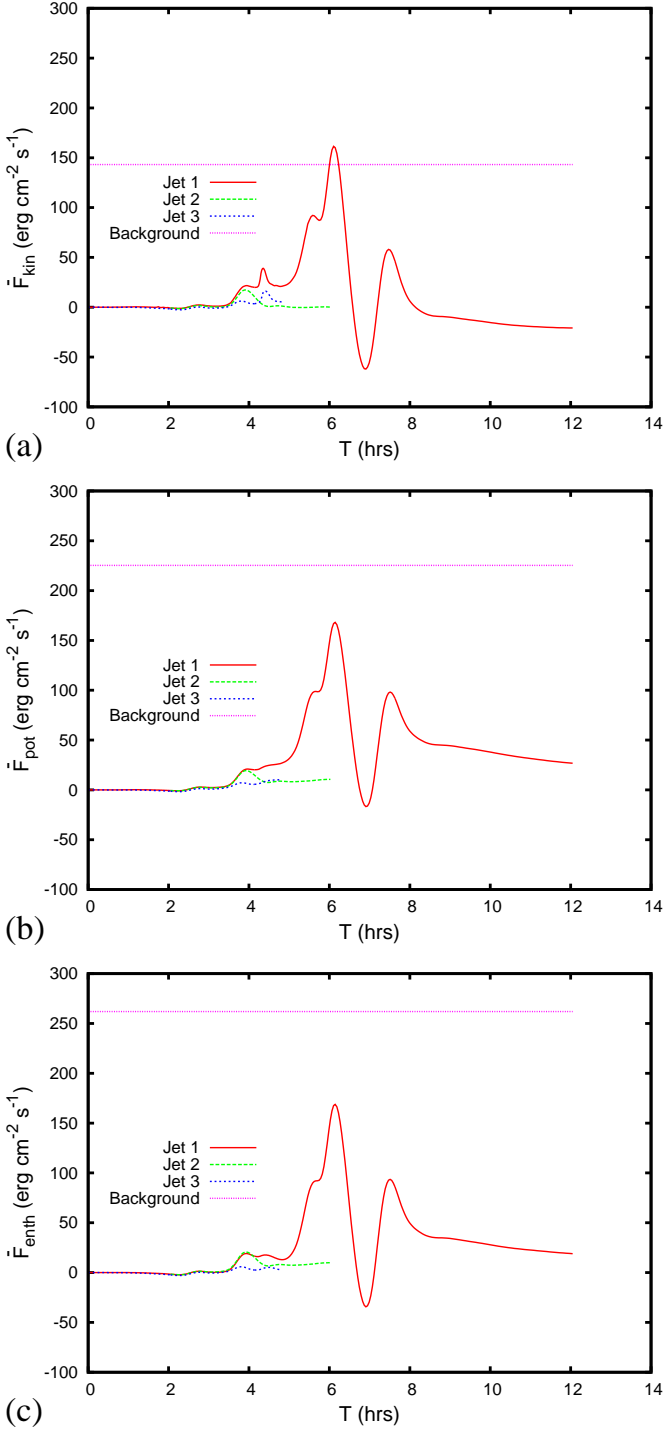}
\caption
{
(a)--(c): Background-subtracted average kinetic-energy, 
potential-energy, and enthalpy flux at $r= 4~R_\odot$ for 
{\em Jet 1--3}. 
The magenta line shows the respective fluxes of the 
background solar wind. 
}
\label{fig-f1_f2}
\end{figure}

\subsection{Wave-Energy Flux vs. Poynting Flux}
\label{ss:wave_vs_poynting}
%
First we compare in Figure~\ref{fig-poynting_wave} the wave-energy flux with the radial Poynting flux. For the former we use the transverse velocities and the magnetic field strength \citep[Equation\,(\ref{eq-flux}); cf. Equation\,(11) in][]{2015A&A...579A..96P}. We see that the wave-energy flux is always smaller than the Poynting flux. In fact, the wave-energy flux formula assumes equipartition of (perpendicular) magnetic and kinetic energy, therefore missing some energy contribution when this condition does not fully apply. Henceforth we use only the Poynting flux to measure the wave energy injected into the SW. 

Prior to the occurrence of the standard jet, no significant wave flux is transferred into the SW. As mentioned above, Alfv\'en waves launched low in the corona require only a few minutes to arrive at $Q$ in our model. Consequently, we see for all runs the first clear signal shortly after the first strong reconnection associated with the onset of the standard jet at $t=1.7$\,h. \emph{Jet 2}, for which the flux emergence is terminated at $t=2$\,h, does not produce any significant signal afterwards. For  {\em Jet\,1} and {\em Jet\,3}, activity increases strongly shortly after $t=3$\,h, marking stronger reconnection episodes associated with the increasing expansion of the emerging flux rope prior to its eruption. The strongest peaks are around $t=4.4$\,h, when signals associated with the blowout jet arrive at $Q$. For {\em Jet\,1}, which was calculated for a longer time than {\em Jet\,3}, the following decay phase contains two secondary peaks. The first one results from small reconnection events in the current sheet that forms below the erupting flux rope, while the second one may be related to a third jet that grows out of the flaring current sheet (T\"or\"ok {\em et al.}, in preparation). The overall evolution of the Poynting flux is very similar for {\em Jet\,1} and {\em Jet\,3}, suggesting a negligible influence of ohmic heating on the strength of the Alfv\'en waves generated by reconnection. A clear deviation can be seen only during the strongest peak, which is associated also with the flux-rope eruption.

\subsection{Evolution of Other Energy Fluxes}
%
Figure\,\ref{fig-f1_f2} shows $\bar{F}_\mathrm{kin}(t)$, $\bar{F}_\mathrm{pot}(t)$, and $\bar{F}_\mathrm{enth}(t)$ together with the background SW flux. \emph{Jet 2} shows relatively weak activity, associated with plasma accelerated during the standard jet, which needs about 2\,hrs to arrive at $Q$. Very similar activity is visible for \emph{Jet 1} but much less for \emph{Jet 3}, which lacks ohmic heating and produces weaker plasma upflows. An additional increase, coinciding with the major peak in the Poynting flux (Figure~\ref{fig-poynting_wave}(b)), occurs in $\bar{F}_\mathrm{kin}(t)$ at $t\,\approx\,4.4$\,h for \emph{Jet 1} and \emph{Jet 3}. Its absence in the other fluxes indicates that it is due to a (transverse) velocity increase rather than a density increase, as the Alfv\'en wave associated with the blowout jet passes through $Q$. 
\p{Indeed, $v_{\perp_\mathrm{max}}=210\,\mathrm{km\,s^{-1}}$ at this time, comparable to the background SW speed ($246\,\mathrm{km\,s^{-1}}$), while $v_r$ and $\rho$ increase by 10 and 30\,\% at most, respectively.}
%

\emph{Jet 1} then shows a strong increase of all fluxes for $t\gtrsim5$\,h, when the blowout jet (i.e., the erupted flux rope) passes through $Q$
%
\p{(which at $t=6.1$ increases $v_r$ and $\rho$ by factors of up to 1.2 and 3.1, respectively).}
\p{The flux maxima ($t=6.1$\,h) are very similar and} 
about one order of magnitude smaller than the peak Poynting flux associated with the preceding Alfv\'en wave (cf. Figure\,\ref{fig-poynting_wave}(b))\p{\footnote{The strong similarity of the flux magnitudes is present only at $r=4\,R_\odot$; their ratios change if the analysis is performed at a different radius.}}.
Afterwards all fluxes show a strong oscillation and a final slow decay. The oscillation has similar amplitude for all fluxes, and is due to the density enhancements and rarefactions visible in Figure\,\ref{fig-jet1}(c). All this activity is barely noticeable in the Poynting flux \p{($v_{\perp_\mathrm{max}}$ = 28\,km\,s$^{-1}$ at $t=6.1$, much smaller than at $t=4.44$).}
For $t\gtrsim8$\,h $\bar{F}_\mathrm{kin}(t)$ remains smaller than $F_\mathrm{kin}^\mathrm{(bg)}$, since $v_r$ falls below the SW speed at $r=4\,R_\odot$ once the ejecta has passed. 

\begin{figure*}
\centering
\includegraphics[width=1.\textwidth]{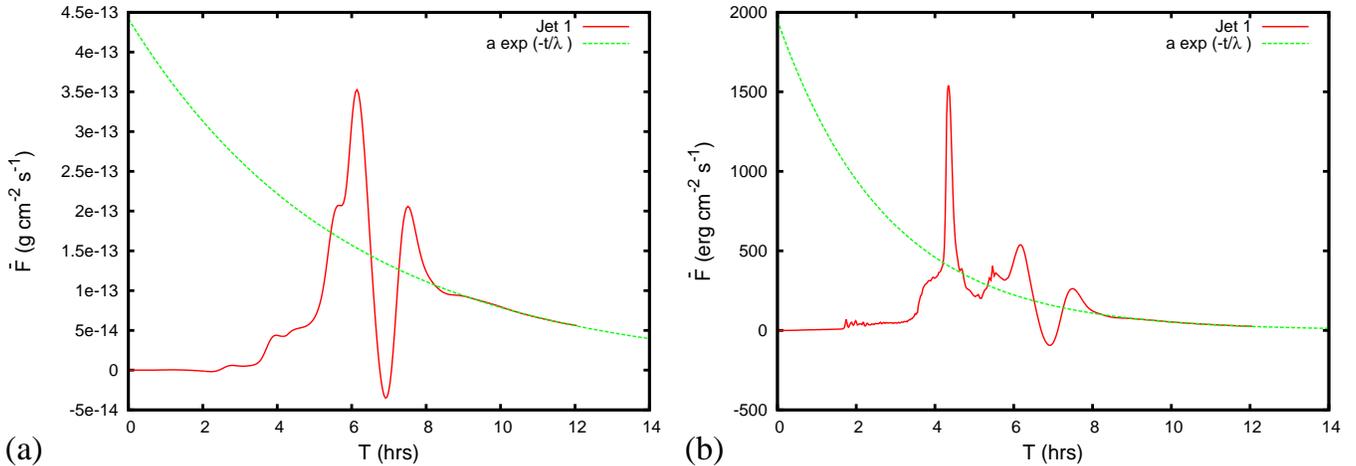}
\caption
{
(a) Mass flux as function of time for run {\em Jet 1}. $\bar{F}$ is extrapolated for $t>12$\,h
using the function $a\,\exp(-t/\lambda)$, with $a=2.83265\times10^{-5}\,\mathrm{g}\,\mathrm{cm}^{-2}\,\mathrm{s}^{-1}$ and $\lambda=20899.6\,\mathrm{s}$.
(b) Same for total-energy flux, with $a=1949.95\,\mathrm{erg}\,\mathrm{cm}^{-2}\,\mathrm{s}^{-1}$,
and $\lambda=9961.97\,\mathrm{s}$.
}
\label{fig-total}
\end{figure*}

\subsection{Total Mass and Energy \& Comparison to Observations}
%
In order to compare our results with observational estimates, we calculate the mass and total energy injected during \emph{Jet 1} into the SW: 
\begin{eqnarray}
M&=& A_\mathrm{jet} \int_0^\infty \bar{F}_\mathrm{mass}\,dt, \nonumber \\
E&=& A_\mathrm{jet} \int_0^\infty  (\bar{F}_\mathrm{kin}+\bar{F}_\mathrm{pot}
+\bar{F}_\mathrm{enth}+ \bar{F}_\mathrm{Poynting})\,dt,
\label{eq-intme}
\end{eqnarray}
where $\bar{F}_\mathrm{mass}$ is the average mass flux calculated from $F_\mathrm{mass}=\rho v_r$ according to Eqs.~(\ref{eq-pt}-\ref{eq-fbar}). Figure\,\ref{fig-total} shows the mass flux and the total energy flux as functions of time. To calculate $M$ and $E$, we extrapolate after $t=12$\,h the integrands to infinity and obtain $M=1.01\times 10^{13}~\mathrm{g}$ and \p{$E=1.57\times 10^{28}~\mathrm{erg}$, the latter corresponding to $\approx$\,1\,\% of the magnetic-energy release during the blowout jet}. $M$ equals the order-of-magnitude estimate obtained from UVCS observations at $1.7\,R_\odot$ by \cite{2007ApJ...659.1702C}. In contrast, both $M$ and $E$ are about one order of magnitude smaller than the mass and kinetic energy estimated by \citet{2014ApJ...784..166Y}. 

Assuming that {\em Jet 1} represents an average coronal jet, a jet-occurrence frequency $f_\mathrm{jet}=(60-240)\,\mathrm{d}^{-1}$ \citep{2007PASJ...59S.771S,2007Sci...318.1580C}, and a fully open magnetic field at $r=4\,R_\odot$, we obtain a total energy flux $F_E=E\,f_\mathrm{jet}/A=\p{(1.1-4.5)}\times10^{1}\,\mathrm{erg}\,\mathrm{cm}^{-2}\,\mathrm{s}^{-1}$ and a proton-density flux $F_M=(M/m_p)\,f_\mathrm{jet}/A=(0.4-1.7)\times10^{10}\,\mathrm{cm}^{-2}\,\mathrm{s}^{-1}$, where $m_p$ is the proton mass and $A$ the spherical surface at $r=4\,R_\odot$. Scaling these fluxes to 1 astronomical unit (AU), assuming spherical SW-expansion at constant speed, we get 
\begin{eqnarray}
F_E&=&\p{(0.4-1.6)}\times10^{-2}\,\mathrm{erg}\,\mathrm{cm}^{-2}\,\mathrm{s}^{-1}\,, \nonumber \\
F_M&=&(1.5-6.0)\times\p{10^{6}}\,\mathrm{cm}^{-2}\,\mathrm{s}^{-1}.
\label{eq:1au}
\end{eqnarray}
$F_M$ \p{is significantly smaller than} the 1\,AU estimate $(1-4)\times10^{7}\,\mathrm{cm}^{-2}\,\mathrm{s}^{-1}$ by \citet{2015ApJ...809..114L} and \cite{2007Sci...318.1580C}\footnote{We use here $f_\mathrm{jet}=(60-240)\,\mathrm{d}^{-1}$ and the correction by \cite{2015ApJ...801..124Y}.}. 

In order to compare with the energy-flux estimate by \cite{2015A&A...579A..96P}, which was obtained for the low corona, we calculate $F_E$ for a typical height extension of coronal jets, $r=1.025\,R_\odot$ \citep{2007PASJ...59S.771S}, and employ the same assumptions for jet frequency and coronal hole area as those authors. We get $F_E = \p{0.21}\times10^{4}\,\mathrm{erg}\,\mathrm{cm}^{-2}\,\mathrm{s}^{-1}$, in \p{very good} agreement with their value $0.18\times 10^{4}\,\mathrm{erg}\,\mathrm{cm}^{-2}\,\mathrm{s}^{-1}$.
  
Finally, we compare our results with SW fluxes derived from in-situ spacecraft measurements. According to \cite{2012SoPh..279..197L}, the total energy flux of the SW at 1\,AU is $\approx$1.5\,erg\,cm$^{-2}$\,s$^{-1}$, and very similar for the slow and fast solar wind. The proton-density flux varies by a factor of about 2 between the slow and fast wind, and falls \p{into} the range $(2-4)\times10^{8}\,\mathrm{cm}^{-2}\,\mathrm{s}^{-1}$ \citep[][and references therein]{2000eaa..bookE2301S,2015ApJ...809..114L}. Comparing these values to our jet-produced fluxes scaled to 1\,AU (Equation\,\ref{eq:1au}), we estimate that coronal jets provide \p{$(0.3-1.0)$}\,\% and \p{$(0.4-3.0)$}\,\% to the total energy and mass content of the SW, respectively.

\section{DISCUSSION AND CONCLUSION}
\label{s:dis}
%
We used an MHD simulation with an advanced energy equation and a large spherical grid to estimate the mass and energy contributions of coronal jets to the SW. The simulation ({\em Jet 1}) produced both a standard and a blowout jet and included ohmic heating. We followed the previous, observational analysis by \cite{2015A&A...579A..96P}, except that we calculated mass and energy fluxes at $r=4~R_\odot$ rather than in the low corona, and used the Poynting flux instead of the wave-energy flux to obtain the total energy flux. Our results can be summarized as follows.

(1) The blowout jet produces a strong perturbation in the SW, the standard jet only to a minor one. This is not surprising, as the former is much more energetic than the latter in our simulation (the peak kinetic energies differ by almost two orders of magnitude). Such a large difference is in line with the results of \citet{2013ApJ...776...16P}, who estimated the magnetic energy released during a blowout jet to be about 20--30 times larger than for a standard jet. 

(2) The blowout jet (and associated flux-rope eruption) triggers a strong Alfv\'en wave that arrives at $r=4\,R_\odot$ more than one hour before the ejecta. This significant lag is due to the relatively large coronal Alfv\'en speed in our model, and is expected to decrease if more realistic values for the Alfv\'en speed in coronal holes are chosen. The wave yields the dominant contribution to the  Poynting flux, while the contributions of Alfv\'en waves triggered by the standard jet and by interim reconnection events remain relatively small. 

(3) The bulk kinetic energy, potential energy, and enthalpy fluxes are also dominated by the blowout jet. In contrast to the Poynting flux, the evolution of these fluxes is characterized by a pronounced oscillation, which reflects plasma density enhancements and rarefactions associated with the eruption. 

(4) The mass and total energy contributions of the blowout jet (the contributions by the standard jet are negligible) at $r=4\,R_\odot$ are $M=1.01\times10^{13}\,\mathrm{g}$ and \p{$E=1.57\times 10^{28}\,\mathrm{erg}$}, respectively. These are only $\sim\,1/10$ of the heliospheric mass and kinetic energy estimates by \citet{2014ApJ...784..166Y}, but those authors considered three of the largest jets during a three-week {\em Hinode} campaign 
\citep[][]{2013ApJ...775...22S}. On the other hand, $M$ agrees with the order-of-magnitude estimate by \cite{2007ApJ...659.1702C} obtained at $r=1.7\,R_\odot$.

(5) Using observed occurrence-rates of coronal jets, we estimated ranges for the total energy and proton-density fluxes produced by these events \p{in the low corona and at 1\,AU. We found very good agreement with the low-corona energy-flux estimate by \cite{2015A&A...579A..96P}. Our estimate for the proton-density flux at 1\,AU is almost an order of magnitude smaller than the values given by \cite{2007Sci...318.1580C}, \citet{2015ApJ...809..114L}, but it must be kept in mind that their estimate was obtained merely from {\em Hinode}/XRT observations.} 

(6) Comparing our results with in-situ measurements of SW fluxes, we estimate that coronal jets provide \p{$(0.3-1.0)$}\,\% and \p{$(0.4-3.0)$}\,\% to the total energy and mass content of the SW, respectively.

Our results are largely consistent with existing observational estimates of the contribution of coronal jets to the mass and energy content of the SW, and thus support the present conjecture that the \p{these contribution are relatively small.}
However, it must be kept in mind that (i) obtaining such estimates requires simplifying assumptions, (ii) the measurement uncertainties are large, and (iii) our simulation is relatively idealized and only one parameter-set \p{of the model (assumed to be representative)} was considered \p{so far}. More advanced observations (for instance from the upcoming Solar Orbiter and Solar Probe Plus missions) and simulations are needed to finally pin down the contribution of coronal jets to the SW.
\acknowledgments
We thank the referee for helpful comments and acknowledge support by NASA's H-SR and LWS programs. M.G.L. and J.E.L. were is part supported by the Chief of Naval Research. This work benefited from discussions within the ISSI International Team 258 ``Understanding Solar Jets and their Role in Atmospheric Structure and Dynamics''.


\end{document}